\documentclass[%
 reprint,
superscriptaddress,
showpacs,preprintnumbers,
 amsmath,amssymb,
 aps,
prb,
]{revtex4-1}

\usepackage{comment,color}

\usepackage{graphicx}
\usepackage{dcolumn}
\usepackage{bm}

\begin{document}


\title{Dispersion of Fermi arcs in Weyl semimetals and their evolutions to Dirac cones
}

\author{Ryo Okugawa}
\affiliation{%
 Department of Physics, Tokyo Institute of Technology, 2-12-1 Ookayama, Meguro-ku, Tokyo 152-8551, Japan
}%
\author{Shuichi Murakami}%
\affiliation{%
 Department of Physics, Tokyo Institute of Technology, 2-12-1 Ookayama, Meguro-ku, Tokyo 152-8551, Japan
}%
\affiliation{%
 TIES, Tokyo Institute of Technology, 2-12-1 Ookayama, Meguro-ku, Tokyo 152-8551, Japan
}%

\date{\today}

\begin{abstract}
We study dispersions of Fermi arcs in the Weyl semimetal phase by constructing an effective model. 
We calculate how the surface Fermi-arc dispersions for the top- and bottom surfaces merge into the bulk Dirac cones in the Weyl semimetal 
at both ends of the arcs, and 
show that they have opposite velocities. 
{This result is common to general Weyl semimetals, and} is also confirmed by a calculation using a tight-binding model. Furthermore, by changing a 
parameter in the system while preserving time-reversal symmetry, we show that two Fermi arcs evolve 
into a surface Dirac cone when the system transits from the Weyl semimetal to the topological insulator phase.  
We also demonstrate that choices of surface terminations  affect the pairing of Weyl nodes, from which the Fermi arcs are
formed.
\end{abstract}

\pacs{73.20.At, 73.43.Nq, 72.25.Dc}

\maketitle

\section{INTRODUCTION}
Topological classification of phases has been one of the fruitful ways to explore new 
quantum phases in condensed materials.
A topological insulator (TI) is one of the topological phases in condensed materials with time-reversal (TR) symmetry
\cite{Kane05b,bernevig2006quantum}. 
As a manifestation of topological properties of three-dimensional TIs, their band structure is gapped in the bulk, but is gapless in the surface.
The surface states are determined by topological invariants calculated from the bulk states.
The topological invariants are Z$_2$ topological numbers defined in TR invariant systems, and 
the resulting gapless surface states are protected topologically.

On the other hand, more recent works have revealed another kind of topological phases, 
not in insulators but in semimetals: for example, Weyl semimetals (WSMs).
In WSMs, the conduction and valence bands form nondegenerate Dirac cones.
They touch at isolated points in $\bm{k}$ space, called Weyl nodes.
Remarkably, the Weyl nodes are stable topologically, because they are associated with a topological number 
called a monopole charge in $\bm{k}$ space, associated with the Berry curvature in $\bm{k}$ space.
The topological WSM phases are realized in 3D systems where TR or inversion (I) symmetry is broken.
As candidates of the WSMs with broken TR symmetry, 
pyrochlore iridates \cite{Wan,Yang},
multilayer structures consisting of TI with ferromagnetic order and normal insulator (NI) \cite{BurkovPRL}, and
HgCr$_2$Se$_4$\cite{Xu}
are proposed.
The multilayer structure of TI and NI with an external electric field is also suggested 
as a candidate material for the WSM where I symmetry is broken. \cite{Halasz}
While the WSMs have not been found experimentally yet, materials with 3D doubly-degenerate Dirac cones were recently discovered, which are called Dirac semimetals.
\cite{Borisenko-Cd3As2,liu2013discovery,neupane2014observation,liu2014stable}
Because of the degeneracy, the gapless points of the Dirac semimetals are not protected topologically, unlike those of WSMs.

The number of Weyl nodes in 3D $\bm{k}$ space is necessarily even. It is
because the Weyl nodes are either a monopole or an antimonopole in $\bm{k}$ space  \cite{Berry84,Volovik,Murakami07b}, and the sum of the monopole charge 
inside the Brillouin zone should vanish.
Moreover, these Weyl nodes are robust topologically as long as translational symmetry is preserved.

Another remarkable topological property of WSMs is the existence of topologically protected surface states \cite{Wan}. 
The surface states form arcs in $\bm{k}$ space, which are called Fermi arcs, connecting between the Weyl nodes projected to the surface Brillouin zone.
The appearance of Fermi arcs is explained in terms of a topological number when the Fermi energy is exactly on the Weyl nodes. 
As a result, the Fermi arc connects two Weyl nodes, one being a monopole and the other an antimonopole for the Berry curvature.
On the other hand, the dispersion of the Weyl nodes is not well studied 
when the Fermi energy is away from the Weyl nodes.
It is easily seen that the sign of the velocity of the 
Fermi-arc state is determined from the topological number, i.e. the monopole charge of the Weyl nodes, 
Moreover, as we find in this paper, the Fermi-arc dispersion has a unique form, which is useful to experimentally 
establish the WSM phase.  
In this paper, we discuss surface state dispersion and bulk bands by constructing a simple effective model for the
WSM phase. The results from the effective model are confirmed by numerical calculations using a lattice model, which is 
the Fu-Kane-Mele model with an additional staggered on-site potential.


{In the present paper, we discuss dispersion of surface states of the WSM phase, and their evolutions at 
phase transitions from the WSM phase to other bulk insulating phases such as TI phases.
In particular, we focus on systems without I symmetry where WSM and TI phases are realizable.
Firstly, from the effective model, we show surface Fermi arcs and their dispersions on the top surface 
and on the bottom surface.
The top- and bottom-surface states have opposite velocities, and are tangential to bulk Dirac cones.
Next, by using a lattice model realizing TI and WSM phases, we study changes of the surface states.
As a result, we find that a pair of Fermi arcs evolve into a surface Dirac cone when the system moves from the WSM to the TI phase.
Furthermore, the pairing of the Weyl nodes to form the Fermi arc depends on the surface termination.
We also discuss how these results are applied to general WSMs. 
}


\section{Weyl semimetal phase characterized by the Berry curvature\label{sec2}}
To characterize WSMs, the Berry curvature in $\bm{k}$ space
is important. As we explain below, the Weyl nodes are topological objects, 
corresponding to monopoles 
or antimonopoles for the Berry curvature. This gives a strong restriction on 
behaviors of Weyl nodes. 
The Berry curvature in $\bm{k}$-space is introduced as follows. 
\cite{Berry84,Volovik,Murakami07b}. 
Let $\psi_{\alpha}(\bm{k})$ be the Bloch wavefunction and 
we write $\psi_{\alpha}(\bm{k})=u_{\alpha}(\bm{k})e^{i\bm{k}\cdot\bm{r}}$, and
$u_{\alpha}(\bm{k})$ is called a periodic part of the Bloch wavefunction. 
For the $\alpha$-th band, the Berry connection (gauge field) $\bm{A}_{\alpha}
(\bm{k})$ and the 
corresponding Berry curvature (field strength) $\bm{B}_{\alpha}(\bm{k})$ are defined as
\begin{align}
&\bm{A}_{\alpha}
(\bm{k})=i\langle u_{\alpha}(\bm{k})|\nabla_{\bm{k}}|
u_{\alpha}(\bm{k})\rangle,\label{eq:A}\\
&\bm{B}_{\alpha}
(\bm{k})=\nabla_{\bm{k}}\times\bm{A}_{\alpha}(\bm{k}),
\label{eq:B}
\end{align}
and the corresponding monopole density is defined as 
\begin{equation}
\rho_{\alpha}
(\bm{k})=\frac{1}{2\pi}\nabla_{\bm{k}}\cdot\bm{B}_{\alpha}
(\bm{k}).
\end{equation}

The properties of the monopole density are well studied \cite{Berry84,Volovik,Murakami07b}.
Therefore we briefly outline its properties here.
When the $\alpha$-th band is not degenerate with other bands at some $\bm{k}$, 
the monopole density $\rho_{\alpha}
(\bm{k})$ vanishes identically, because 
$u_{\alpha}(\bm{k})$ is analytic in the
neighborhood of $\bm{k}$.
Only at the $\bm{k}$-points where the $\alpha$-th band 
touches with another band, 
the monopole density can be nonzero, having 
a $\delta$-function singularity. 
It is shown from an argument on a gauge degree of freedom that the coefficient of
$\delta$-function is always an integer, and the monopole density has the form $\rho(\bm{k})=\sum_{l}q_{l}
\delta(\bm{k}-\bm{k}_{l})$, where $q_l$ is an integer. We call the integer $q_{l}$ 
a monopole charge.
In the simplest case of $q_l=\pm 1$ is called a monopole ($q_{l}=1$) and
an antimonopole ($q_{l}=-1$). The Weyl nodes are either a monopole or an antimonopole, as one can see easily from 
an example Hamiltonian $H=\bm{k}\cdot \bm{\sigma}$, where $\bm{\sigma}=(\sigma_x,\sigma_y,\sigma_z)$
are the Pauli matrices. 
Because the monopole charge is quantized, the monopole charge cannot change under a
continuous change of the Hamiltonian. They can only change at pair creation or pair annhilation of
a monopole-antimonopole pair. 
 
TR and I symmetries respectively give a restriction to these Berry curvature and 
monople density. 
The TR symmetry leads to 
\begin{equation}
\bm{B}_{\alpha}(\bm{k})=-\bm{B}_{\alpha_R}(-\bm{k}), \ 
\rho_{\alpha}(\bm{k})=\rho_{{\alpha}_R}(-\bm{k}), 
\label{eq:time-reversal-rho}
\end{equation}
where ${\alpha}_R$ is the band index obtained by time-reversal from $\alpha$th band.
Hence in TR-symmetric systems, monopoles are distributed symmetrically 
with respect to the origin $\bm{k}=0$. 
On the other hand, the I symmetry leads to 
\begin{equation}
\bm{B}_{\alpha}(\bm{k})=\bm{B}_{{\alpha}_I}(-\bm{k}), \ 
\rho_{\alpha}(\bm{k})=-\rho_{{\alpha}_I}(-\bm{k}), 
\end{equation}
where ${\alpha}_I$ is the band index obtained by inversion from $\alpha$th band.
Hence in I-symmetric systems, monopoles are distributed antisymmetrically 
with respect to the origin $\bm{k}=0$. 
Furthermore, in systems with both TR and
I symmetries, all states are doubly degenerate by Kramers theorem, and therefore
a Dirac cone without degeneracy is impossible. 
Therefore, the WSM requires either breaking of TR symmetry 
or that of I symmetry, as has been proposed \cite{BurkovPRL,BurkovPRB}.
Such systems with broken TR or I symmetries can be realized as multilayers of TIs and NIs \cite{BurkovPRL,BurkovPRB,Halasz}.

\section{Effective model for the NI-SW-TI phase transition\label{sec3}} 
To describe a WSM and its evolution under a change of Hamiltonian parameters, 
we construct a minimal model including only a single valence band and a 
single conduction band.
Therefore, we consider a minimal model 
described by a $2\times 2$ matrix $H(\bm{k},m)$, where
$m$ is introduced as a control parameter for NI-WSM-TI phase 
transition.
The 2$\times$2  Hamiltonian $H(\bm{k},m)$ is expanded as
\begin{equation}
H(\bm{k},m)=a_0(\bm{k},m)+\sum_{i=1}^{3}a_{i}(\bm{k},m)
\sigma_i,
\label{eq:hamiltonian}
\end{equation}
where $\sigma_i$ ($i=1,2,3$) are the Pauli matrices representing conduction and valence bands. 
The gap between the two eigenvalues closes 
when the three conditions
\begin{equation}
a_{i}(\bm{k},m)=0 \ \ (i=1,2,3)
\label{eq:gaplesscondition}
\end{equation}
are satisfied. 
If Eq.~(\ref{eq:gaplesscondition}) has solutions 
for $\bm{k}$ at a given value of $m$, the bands generally form a
Dirac cone without degeneracy at these $\bm{k}$ points, if $\frac{\partial(a_1,a_2,a_3)}{\partial(k_x,k_y,k_z)}\neq 0$. 
Therefore it is generally a WSM and 
the respective Weyl nodes are monopoles or antimonopoles,
depending on the monopole charge equal to ${\rm sgn}\frac{\partial(a_1,a_2,a_3)}{\partial(k_x,k_y,k_z)}=\pm 1$.

Let us then change the parameter $m$. In order to open a gap in the system, 
all the monopoles and antimonopoles should disappear via monopole-antimonopole pair annihilation. 
Conversely, if we begin with a system with a bulk gap at some parameter $m$, and if
a change of $m$ induces appearance of Weyl nodes, then it should involve
monopole-antimonopole pair creation. One of the purposes of the present paper is to create a simple
effective model describing the WSM phase close to
monopole-antimonopole pair creation or annihilation, i.e. near the phase transition 
between the WSM phase and a bulk insulating phase. Let $m=m_0$ be
the value of $m$ where this monopole-antimonopole pair creation occurs.
A part of the formalism here follows the previous paper by one of the authors \cite{MurakamiKuga}. 

Suppose we change the value of $m$ through the phase transition between the WSM phase and a
phase with a bulk gap. It is accompanied by a pair creation or annihilation, and let $m_0$ 
denote the value of $m$ where the phase transition occurs. Then on one side of $m$, e.g. $m<m_0$
the system is a WSM with an monopole and antimonopole, 
while on the other side of $m$, e.g. $m>m_0$ the system is an insulator in the bulk, which 
can be a NI or a TI. At $m=m_0$ the gap closes at some point where the pair creation occurs, 
and let 
$\bm{k}=\bm{k}_{0}$ denote the point; namely, it satisfies 
$\bm{a}(\bm{k}_{0},m_{0})=0$.
We expand the coefficients of Eq.~(\ref{eq:hamiltonian}) to the linear order around $\bm{k}_{0}$ and $m_{0}$:
\begin{equation}
\bm{a}(\bm{k},m)=
M\Delta \bm{k}
+\Delta m\bm{N},
\label{eq:a-matrix}
\end{equation}
where $\Delta k_{j}=k_{j}-k_{0j}$, $\Delta m=m-m_{0}$, and 
$M_{ij}=\left.\frac{\partial a_{i}}{\partial k_{j}}
\right|_{0}$ and 
$N_i
=\left.\frac{\partial a_{i}}{\partial m}\right|_{0}$ are the values of derivatives at $m=m_0$ and $\bm{k}=\bm{k}_0$.
It gives a generic Hamiltonian describing a pair creation of monopoles 
at $m=m_0$, shown in Ref.~\onlinecite{MurakamiKuga}. 
From this Hamiltonian we can calculate the motion of the Weyl nodes close to the pair creation
(i.e. the phase transition between the WSM and the bulk insulating phase), and band dispersions
\cite{MurakamiKuga}. 

It is also noted in Ref.~\onlinecite{MurakamiKuga} that pair creations occur in pairs
at $\bm{k}=\bm{k}_0$ and  $\bm{k}=-\bm{k}_0$ simultaneously,
imposed by Eq.~(\ref{eq:time-reversal-rho}). Thus there are at least two monopoles and two antimonopoles
in the WSM with TR symmetry (but without I symmetry). 
If one varies $m$ further and the system becomes a bulk insulating phase 
again, there should be pair annihilations to make all the monopoles and antimonopoles disappear. 
If pair annihilations occur by exchanging partners from the pair creations, some of the four 
$Z_2$ topological numbers of 3D TIs should be different between the initial bulk-insulating phase  
and the final bulk-insulating phase. 

Our goal here is to calculate an evolution of surface states through this change of the WSM phase. 
To this goal the generic Hamiltonian described above is not convenient because it contains  many 
parameters. Therefore, instead of using the above generic Hamiltonian, we
use a simplified Hamiltonian. This is obtained from the above Hamiltonian after
some gauge transformation, scale transformation, and a few simplifying assumptions. 
The details of this derivation is in Appendix \ref{sec:app-A}.
The resulting effective model is described by a Hamiltonian
\begin{equation}
H=\gamma(k_x^2-m)\sigma_x+v(k_y\sigma_y +k_z\sigma_z),
\label{eq:H}
\end{equation}
where $v$ and $\gamma$ are nonzero constants, and we choose them to be positive
for simplicity.
{In deriving this model, among Weyl nodes in the WSM, we focused on one monopole and one antimonopole
that are assumed to be close to each other. 
We then shifted the origin of the wavevector to simplify the Hamiltonian and retained terms which are of the lowest order in $\bm{k}$. 
Therefore, this Hamiltonian generally 
applies to any WSMs, i.e. those with I symmetry breaking or those with 
TR symmetry breaking, as long as a monopole and an antimonopole are close to each other.
We note that the origin of the wavevector in this Hamiltonian does not correspond to $\bm{k}=0$ in the original Bloch wavevector due to the shifting.
Therefore symmetry properties of 
the Hamiltonian $H$, such as I or TR symmetries, cannot be discussed in Eq.~(\ref{eq:H}).
}
Its bulk dispersion is given by 
\begin{equation}
E=\pm\sqrt{\gamma^2(k_x^2-m)^2+v^2k_y^2+v^2k_z^2}.
\label{eq:bulk}
\end{equation}
The Fermi energy is assumed to be at $E=0$.  
When $m<0$ it describes a phase with a bulk gap $2\gamma|m|$, either the TI or the NI phase. 
On the other hand, when $m>0$ the bulk gap closes at two points 
W$_\pm$: $\bm{k}=(\pm\sqrt{m}, 0,0)$. 
Around these points the dispersions are linear in three directions, and 
therefore they describe the WSM phase. These two Weyl nodes W$_+$ and W$_-$ are a 
monopole and an antimonopole for the lower band, respectively. At $m=0$ there occurs a monopole-antimonopole 
pair creation. 
{Hence, this effective Hamiltonian describes one pair creation/annihilation of a monopole and an antimonopole when the phase transition occurs.}

Let us consider a surface of the WSM phase with $m=m_0(>0)$. 
Following the standard technique, we describe the surface by a space-dependent value of $m$.
Namely, we regard $m$ to change its 
sign at the surface. We set $m$ to have a negative value in the vacuum side, and $m$ converges to $m_0(>0)$ in 
the WSM side.
The surface is assumed to be along the $xy$ plane for simplicity. 
Therefore we set
\begin{equation}
\left\{
\begin{array}{ll}
m(z)=m_0 &:\ z\rightarrow -\infty\\
m(z)<0 &:\ z\rightarrow +\infty
\end{array}\right.
\end{equation}
for the surface normal to be $+\hat{z}$, which we call a top surface, and 
\begin{equation}
\left\{\begin{array}{ll}
m(z)<0 &:\ z\rightarrow -\infty\\
m(z)=m_0 &:\ z\rightarrow +\infty
\end{array}\right.
\end{equation}
for the surface normal to be $-\hat{z}$, which we call a bottom surface.
They correspond to the top and bottom surfaces of a slab with sufficiently large thickness.

Here we calculate band dispersions for the top surface and for the bottom surface. 
By unitary transformation with $U=\frac{1}{\sqrt{2}}(1-i\sigma_x)$, 
the Hamiltonian is transformed to 
\begin{eqnarray}
&&H'\equiv U^{-1}HU=\gamma(k_x^2-m)\sigma_x-iv\frac{\partial}{\partial z}k_z\sigma_y -vk_y\sigma_z\nonumber \\
&&=
\left(\begin{array}{cc} -vk_y& \gamma(k_x^2-m)-v\frac{\partial}{\partial z} \\ \gamma(k_x^2-m)+v\frac{\partial}{\partial z} & vk_y
\end{array}\right)
\end{eqnarray}
Because we focus on the surface within the $xy$ plane, we have replaced $k_z$ with $-i\frac{\partial}{\partial z}
$, while $k_x$ and $k_y$ are the Bloch wavenumbers.
It is now straightforward to write down the eigenstates bound to the surfaces. The bound state on the top surface 
is given by 
\begin{equation}
\psi_{\rm T}=\left(\begin{array}{c}1 \\ 0\end{array}\right)e^{-(\gamma/v)\int^{z}(k_x^2-m(z))dz},\ E=-vk_y
\label{eq:psiT}
\end{equation}
and the bound state on the bottom surface 
is given by 
\begin{equation}
\psi_{\rm B}=\left(\begin{array}{c}0\\ 1\end{array}\right)e^{(\gamma/v)\int^{z}(k_x^2-m(z))dz},\ E=vk_y
\label{eq:psiB}
\end{equation}
They are respectively allocated as top- and bottom-surface states, because otherwise the wavefunction 
diverges at some region and is not normalizable. We also note that both of these surface states exist 
only when 
\begin{equation}
-\sqrt{m_0}<k_x<\sqrt{m_0}.
\label{eq:arc-boundary}
\end{equation}
At $E=0$, the surface states are degenerate, and are located at $k_y=0, -\sqrt{m_0}<k_x<\sqrt{m_0}$, which is a line connecting the 2D projection of the
Weyl 
points ${\rm W}_{\pm}$: $\bm{k}=(\pm\sqrt{m_0},0,0)$.
Thus these surface states 
are Fermi arcs. 
We note that the top-surface Fermi-arc states have a velocity $\bm{v}=\frac{\partial E}{\partial \bm{k}}=(0,-v)$
and those of the bottom-surface  have a velocity $\bm{v}=\frac{\partial E}{\partial \bm{k}}=(0,v)$.
Their signs are consistent with the fact that W$_{\pm}$ are a monopole and an antimonopole, respectively. 
The velocity signs follow from the fact that 
on the slice of the 3D  BZ at $k_x=\rm{const.}$,
{the lower band has a Chern number 0 for  $\sqrt{m_0}<|k_x|$ 
and $-1$ for  $-\sqrt{m_0}<k_x<\sqrt{m_0}$
due to the antimonopole at W$_{1-}$. }

\begin{figure}
\includegraphics[width=8.5cm]{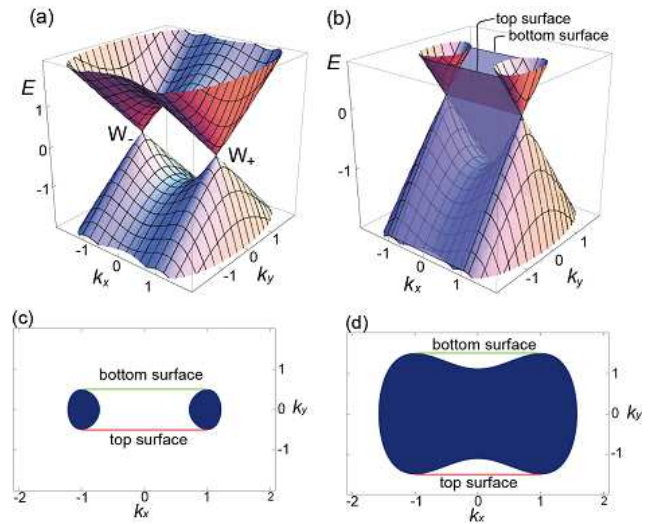}
\caption{\label{pzfill}
Bulk and surface states of the model for the WSM. The parameters 
are set as $v=\gamma =m_0=1$. 
(a) shows the bulk states, and (b) shows the surface Fermi arcs and the bulk states.
(c) and (d) The Fermi surface for the Fermi energy at (c) $E =0.5$ and at (d) $E=1.5$.
}
\end{figure}

We show how these surface states disperse if the Fermi energy is away from the Weyl point. 
To see how the bulk bands and 
surface bands are related, we project the bulk dispersion, 
Eq.~(\ref{eq:bulk}), onto the surface. The resulting bulk bands are in regions
\begin{eqnarray}
&&E>\sqrt{\gamma^2(k_x^2-m_0)^2+v^2k_y^2}, \\
&&E<-\sqrt{\gamma^2(k_x^2-m_0)^2+v^2k_y^2},
\label{eq:bulk-projected}
\end{eqnarray}
which describe the conduction and the valence bands, respectively.
These two bands touch each other at the projections of the two Weyl nodes W$_{\pm}$: $k_x=\pm\sqrt{m_0}$, $k_y=0$.
Around these Weyl nodes the dispersion is linear. 
This projected bulk band structure is shown in Fig.~\ref{pzfill} (a), forming two Dirac cones around the 
Weyl nodes W$_{\pm}$. The surface states are tangential to these cones, as shown in Fig.~\ref{pzfill} (b).
{A similar surface-state dispersion was proposed in Ref.~\onlinecite{Wan} without calculations, and
our calculation on the effective model confirms this dispersion of surface states.
We also note that similar results have been independently obtained 
in Ref.~\onlinecite{Haldane} for a toy model. Our results are based on 
generic considerations (see Appendix) from an effective model (\ref{eq:H}), and are applicable to general WSM.} 

To see the relationship between the projection of the bulk bands and the surface states, 
we show the Fermi surface at a constant energy $E$. The bulk-band projection 
changes its topology at $E=\pm \gamma m_0$. When $-\gamma m_0<E<\gamma m_0$, the bulk-band projection 
forms two distinct pockets as shown in 
Fig.~\ref{pzfill} (c), and the Fermi arcs are bridged between these two pockets. When $|E|>\gamma m_0$, it forms one pocket with 
a dumb-bell-like structure [Fig.~\ref{pzfill} (d)].
In either case, it is remarkable that the Fermi arc merges to the bulk-band projection at the two 
ends, and at both ends they are tangential to the bulk-band projection. 

{We note that the obtained dispersion of surface and bulk states are generic
as long as the monopole and the antimonopole are close to each other. It is because the Hamiltonian is derived from generic consideration by expanding the Hamiltonian in terms of the wavevector close to the monopole-antimonopole pair creation point, with scale transformation. Hence we expect that this dispersion holds for general WSMs with a pair of Weyl nodes that are close to each other. }



\section{NUMERICAL CALCULATION OF SURFACE STATES OF WEYL SEMIMETALS IN A LATTICE MODEL \label{mod}}
\subsection{Model}
In this section, we numerically calculate surface states in a WSM phase and compare the results to the discussions in Sec.~\ref{sec3}.
For this purpose, we begin with the Fu-Kane-Mele (FKM) tight-binding model \cite{Fu06b}, which 
is known to show various 3D TI phases. It does not show the WSM phase as it is, because it does not
break I symmetry. 
By adding a staggered on-site potential to the model to break I symmetry, it does show 
the WSM phase as shown in Ref.~\onlinecite{MurakamiKuga}. It was later used also in Ref.~\onlinecite{Ojanen} to calculate 
surface Fermi arcs in some parameter range. 

Hence, we use the FKM model with a staggered on-site potential added. Our model is 
a 3D tight-binding model on a diamond lattice, described by the following Hamiltonian 
\begin{eqnarray}
H= 
\sum_{\langle i,j \rangle}t_{ij}c_i^{\dag}c_j +
i\frac{8\lambda _{so}}{a^2}\sum_{\langle \langle i,j \rangle \rangle}
c_i^{\dag} \bm{s} \cdot (\bm{d} _{ij}^1 \times \bm{d} _{ij}^2) c_j \nonumber \\
+\lambda _{v}\sum_{i}\xi _i c_i^{\dag}c_i, \label{modelH}
\end{eqnarray}
where $\bm{s}$ are Pauli matrices and $a$ is the lattice constant for the cubic unit cell.
The first term is the nearest-neighbor hopping with hopping amplitude $t_{ij}$.
The second term represents the spin-orbit interaction for next nearest neighbor 
hopping
with a spin-orbit coupling parameter $\lambda _{so}$.
$\bm{d}^1_{ij}$ and $\bm{d}^2_{ij}$ are the nearest neighbor vectors connecting second-neighbor sites $i$ and $j$.
The third term represents the 
staggered on-site energy $\pm \lambda_v$, where $\lambda _v$ is a constant and 
$\xi _i = \pm 1$ depends on the sublattices, i.e., $\xi=+1$ for the A sublattice and 
$\xi=-1$ for B sublattice for the diamond lattice.

The model without the third term is the FKM model, and is TR and I symmetric \cite{Fu06b}.
Provided the nearest-neighbor hoppings $t_{ij}$ are identical, the FKM model has gapless band structure
with the bulk gap closed at the three X points, showing that it is a Dirac semimetal. 
There are four directions of the nearest neighbor bonds, and 
when the hopping integrals for four nearest-neighbor bonds $t_\alpha$ ($\alpha=1,2,3,4$) become different, 
the model shows various phases of either strong topological insulator (STI) or weak topological insulator (WTI) phases.
The hopping integrals along the bond in the 111, 1$\bar{1}\bar{1}$, $\bar{1}1\bar{1}$, and $\bar{1}\bar{1}1$ directions are denoted as $t_1$, $t_2$, $t_3$, and $t_4$, respectively.  

To realize the WSM phase with TR symmetry, the system needs to be I-asymmetric.
In Ref.~\onlinecite{MurakamiKuga}, it is shown that with the $\lambda _v$ term breaking the I-symmetry, this model shows the WSM phase.
The Hamiltonian matrix is 
\begin{align}
\mathcal{H}(\bm{k})&= \begin{pmatrix}
			\lambda _v \bm{1}+\sum_{i=1}^3 F_is_i & f\bm{1} \\
			f^{\ast }\bm{1} & -\lambda _v \bm{1}-\sum_{i=1}^3 F_is_i 
			\end{pmatrix}
\end{align}
where
\begin{align}
f&=t_1+t_2e^{i\bm{k}\cdot \bm{a_2}}+t_3e^{i\bm{k}\cdot \bm{a_3}}+t_4e^{i\bm{k}\cdot \bm{a_1}},  \\
F_x&=-4\lambda _{so}\sin \frac{a}{2}k_x \Bigl( \cos \frac{a}{2}k_y -\cos \frac{a}{2}k_z \Bigr) , 
\end{align}
and $F_y$ and $F_z$ are given by cyclic permutation of the subscripts $x$, $y$, and $z$ in $F_x$.
The primitive vectors of the fcc lattice are defined as
$\bm{a}_1=\frac{a}{2}(1,1,0)$, $\bm{a}_2=\frac{a}{2}(0,1,1)$, $\bm{a}_3=\frac{a}{2}(1,0,1)$.
The energy eigenvalues are 
\begin{align}
E(\bm{k})=\pm \sqrt{ (\lambda _v \pm |\bm{F}| ) ^2+|f|^2 },
\end{align}
where $\bm{F}=(F_x,F_y,F_z)$.
Therefore, the spectrum is gapless when
\begin{equation} 
\mathrm{Re} f=\mathrm{Im}f=0,\hspace{2mm} \lambda _{v}=\pm |\bm{F}|. \label{condweyl}
\end{equation}
In some parameter region, the three equations (\ref{condweyl}) for three parameters 
$k_x, k_y$, and $k_z$ have solutions, showing the locations of the Weyl nodes. The bulk gap is then closed and the WSM phase appears there. 

\subsection{Numerical calculation of surface states}
In Ref.~\onlinecite{MurakamiKuga}, phase diagrams of this model are studied and this model is shown to exhibit the STI, WTI, and WSM phases by changing parameters. 
As an example, we assume $t_1=t+\delta t_1$, $t_2=t+\delta t_2$, $t_3=t_4=t$ 
and $\delta t_{-} =\delta t_1 - \delta t_2$ is fixed to be positive while $\delta t_{+} =\delta t_1 + \delta t_2$ is varied.
For the case with I symmetry, i.e. $\lambda_v=0$, a band inversion at $X^x=\frac{2\pi}{a}(1,0,0)$ occurs at $\delta t_+=0$, accompanied
by a phase transition between the STI phase with the $Z_2$ topological number 1;(111) ($\delta t_+>0$) and the WTI phase 
with the $Z_2$ topological number 0;(1$\bar{1}\bar{1}$) ($\delta t_+<0$) \cite{Fu06b}. 
As a result, when the system is in the STI phase, a surface Dirac cone arises at the point 
$X^x=\frac{2\pi}{a}(1,0,0)$ projected onto the surface Brillouin zone,  
while in the WTI phases there is no surface Dirac cone at this point.

\begin{figure}[t]
\includegraphics[height=6cm]{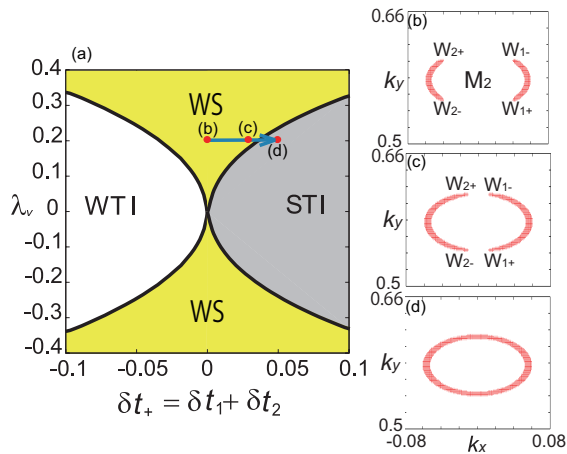}
\caption{\label{conearc}
(a) Phase diagram in the $\delta t_{+}$-$\lambda _{v}$ plane, with $\delta t_{-}=0.1t$ and $\lambda _{so}=0.1t$, 
where $\delta t_{\pm} =\delta t_1 \pm \delta t_2$.
The axes are in the unit of $t$. 
When $\lambda_{v}=0$, the system is I-symmetric, and no WSM phase appears.
(b)-(d) Surface Fermi surface at $E=0$ around the point $M_2$ near the Fermi level with $\lambda _v=0.2t$ and 
the following values for $\delta t_+$:
(b) $\delta t_+=0$ (WSM), (c) $\delta t_+=0.03t$ (WSM), and (d) $\delta t_+=0.05t$ (STI).
The axes are in the unit of $2\pi/b$. 
(b) and (c) show Fermi arcs in the WSM phase, and (d) shows a surface Dirac cone in the STI phase.
We note that the end points of the surface Fermi arcs in the WSM phase are the gapless points of the bulk bands.
}
\end{figure}

If one introduces an on-site staggered potential $\lambda_v$, the I symmetry is broken while the
TR symmetry is preserved. Then the WSM phase intervenes between the STI and the WTI phases,
as shown in Fig.~\ref{conearc} (a)
In the WSM phase, there are four Weyl nodes around the $X^x$ point, as found in Ref.~\onlinecite{MurakamiKuga}. These four Weyl nodes move as the parameter $\delta t_{+} = \delta t_1 + \delta t_2$ changes. Among these four Weyl nodes, two are monopoles and the other two are antimonopoles, 
which distribute symmetrically with respect to the $X^x$ point. 
On the surface, two Fermi arcs will arise, connecting monopole-antimonopole pairs. 
For calculations we fix $\delta t_{-} = \delta t_1-\delta t_2=0.1t$ and $\lambda _{so} =0.1t$.

To see surface states, we numerically diagonalize Eq.~(\ref{modelH}) in a slab geometry 
with (111) surfaces. To show the surface states, we take the $z$ axis to be the surface normal along $\left[ 111\right]$, the $x$ axis along the surface in the $\bm{a_3}-\bm{a_1}$ direction 
and the $y$ axis to be perpendicular to the $x$ and $z$ axes. 
The top surface of the slab is composed of lattice sites in the sublattice A and the bottom surface is composed of lattice sites in the sublattice B.
Because the point $X^x$ is projected to the point $M_2=\frac{2\pi}{b}(0,1/\sqrt{3})$ in the hexagonal surface Brillouin zone,
the Dirac cones and the Fermi arcs appear around the point $M_2$.
Here $b=a/\sqrt{2}$ is the length of the primitive vectors of the slab.

Figures \ref{conearc} (b)-(d) shows Fermi surfaces of a slab at $E=0$ for various values of $\delta t_+$, with 
$\delta t_{-}=0.1t$ and $\lambda _{so}=0.1t$. 
For (b) $\delta t_+=0$ and (c) $\delta t_+=0.03$, the system is 
in the WSM phase, and Fermi arcs appear around point $M_2$, corresponding to point $X^x$. 
The ends of arcs are the Weyl nodes projected into the surface Brillouin zone. 
Among the four Weyl points, let W$_{1+}$ and W$_{2+}$ denote the monopoles, and 
let W$_{1-}$ and W$_{2-}$ denote the antimonopoles, which are shown in 
Figs.~\ref{conearc} (b) and (c). 
As $\delta t_+$ is changed, the Weyl nodes 
move around this $M_2$ points, and concomitantly the Fermi arcs grow
as seen in Figs.~\ref{conearc}(b) and (c).
As $\delta t_+$ is increased further, the system eventually enters the STI phase. At the WSM-STI phase transition, 
the Weyl nodes annihilate pairwise for (W$_{1-}$,W$_{2+}$) and  (W$_{1+}$,W$_{2-}$), and there is no Weyl node in the STI phase, with a nonzero bulk gap. 
Correspondingly, as we see in Fig.~\ref{conearc}(d), the two Fermi arcs in the WSM phase are merged into a 
surface Dirac cone in the STI phase.

We discuss a relationship between the present work and the paper by Ojanen \cite{Ojanen}. 
In the present work we fix the spin-orbit coupling $\lambda_{so}$ to be a constant
and change the anisotropy of the nearest neighbor hoppings. It enables us  the phase transitions 
from the WSM phase to either the STI or the WTI phases. On the other hand, in Ojanen's paper \cite{Ojanen}
 the spin-orbit parameter $\lambda_{so}$ is changed across zero, to see the NI-WSM phase transitions. 
In approaching the WSM-NI phase transition, the Fermi arcs are gradually shortened
 and eventually vanish.

\begin{figure}[t]
\includegraphics[width=7cm]{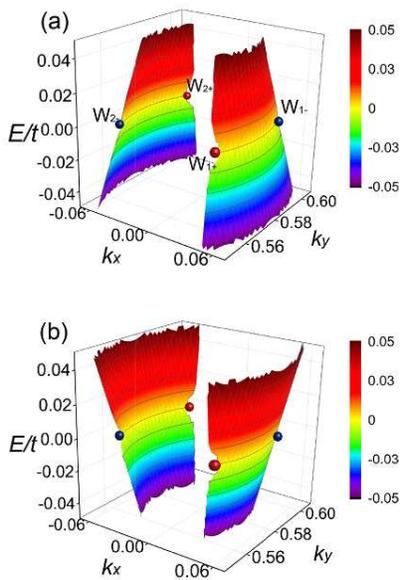}
\caption{\label{arcside}Side views of the Fermi arcs around point $M_2=\frac{2\pi}{b}(0,1/\sqrt{3})$ with $\lambda _v=0.2t$, $\delta t_+=0$ (WSM phase).
for (a) the top surface and (b) the bottom surface.
The $k_x$ and $k_y$ axes are in the unit of $2\pi/b$. 
The red (blue) points are the gapless points which have positive (negative) monopole charges for the Berry curvature.
}
\end{figure}

So far we have discussed the surface states on $E=0$, where the states on the top surface and those on 
the bottom surface are degenerate. 
The top-surface states and bottom-surface states are expected to have different dispersions, as Fig.~\ref{pzfill} (b) 
shows. Figure \ref{arcside} shows the results for the dispersion of the 
Fermi arcs on the top- and bottom-surface states in the present model. 
{We note that the top- and bottom-surface states between a pair of Weyl nodes have opposite velocities,
and the signs of the velocities are consistent with the monopole charges of W$_{i\pm}$ ($i=1,2$). }
{To see this, let us focus on the surface
Fermi arc between W$_{1+}$ and W$_{1-}$ as an example, and ignore the other Fermi arc.
Let us take a 2D slice of the 3D Brillouin zone, which includes the surface normal
($[111]$ direction). If the slice does not intersect the line between W$_{1\pm}$,
the Chern number is zero within this 2D slice, while it is one when the slice intersects the
line between W$_{1\pm}$ because of the presence of the monopole at W$_{1+}$. This means 
that within this slice there should be a clockwise topological edge mode, which appears 
as a surface mode with negative velocity $v_x<0$ on the top surface and 
that with positive velocity $v_x>0$ on the bottom surface.}
As is consistent with the result of
the effective model [Fig.~\ref{pzfill} (b)],
each of these surface Fermi arcs is bridged between two Dirac cones around the Weyl nodes.  
As $\delta t_+$ is increased and the system undergoes the phase transition from the WSM phase into the 
STI phase, the two Fermi arcs merge into a single Dirac cone on the top surface, and the same occurs
on the bottom surface. As a result there arises a top-surface Dirac cone and a bottom-surface Dirac cone, 
which are nondegenerate as shown schematically in Fig.~\ref{arcandcone}(a). This splitting of the 
Dirac cones are natural, because of the breaking of the I-asymmetry due to the staggered 
on-site energy $\lambda _v$. 
In the present case, the topmost layer in the top (bottom) surface is A sublattice (B sublattice),
and therefore the top-surface (bottom-surface) states have a larger (smaller) energy due to
the staggered 
on-site energy $\lambda _v$.

The surface states in the whole BZ for $\delta t_+=0$ (WSM) and $\delta t_+=0.05t$ (STI) when $\lambda _v =0.2t$ are shown in Figs.~\ref{dbnew} (a1) and (b1). In 
addition to the surface states around M$_2$, there are Dirac cones around M$_1$ and M$_3$. Nevertheless, they 
are intact at the WTI-WSM-STI phase transition, because this phase transition is related with a 
band inversion at X$_x$, which is projected onto M$_2$ point. 

\subsection{WSM-TI phase transition and evolution of the Fermi-arc surface states}
Based on the calculation results on the model (\ref{modelH}), 
here we discuss general features of the evolution of the Fermi-arc surface states in the WSM phase when 
some parameter is changed. 
In the WSM phase there are an even number of Weyl nodes. In the I-asymmetric phases with TR symmetry, the minimal number is four, 
i.e., two monopoles and two antimonopoles, as follows from Eq.~(\ref{eq:time-reversal-rho}). In this case of 
two monopoles and antimonopoles, symmetrically distributed around a TRIM $\bm{k}_\Gamma$, the Fermi arcs are
formed between monopole-antimonopole pairs, as exemplified in Fig.~\ref{conearc}. Suppose then we change some parameter in the system. 
Due to topological nature of Weyl nodes, the monopoles and antimonopoles move in the 2D surface BZ. 
Eventually, they may undergo some pair annihilations, which occur symmetrically with respect to the TRIM 
$\bm{k}_\Gamma$; the bulk bands become gapped. 
If a pair annihilation occurs for a pair connected by a Fermi arc, the Fermi arc eventually vanishes 
and the surface becomes gapped. 
On the other hand, if the pair annihilation occurs between a monopole and an antimonopole, which are not 
connected to each other by the Fermi arc, the pair annihilations will make all the Fermi arcs into a single loop
encircling the TRIM $\bm{k}_\Gamma$. This loop constitutes a Dirac cone around the TRIM. 

\begin{figure}[b]
\includegraphics[width=7cm]{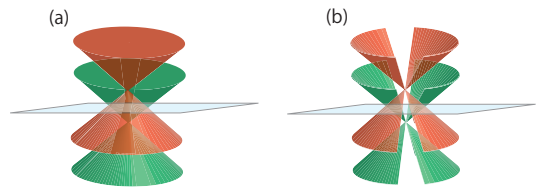}
\caption{\label{arcandcone}Schematic drawing of the surface energy bands around a TRIM. 
The red (green) cone is the top (bottom) surface states.
(a) In the STI phase, the two Dirac cones on the top and bottom surfaces are split in energy when the I symmetry is broken.
(b) In the WSM phase, there are a pair of Fermi arcs on each surface. These Fermi arcs will evolve into a Dirac cone shown in (a) in the STI phase.}
\end{figure}

From the viewpoint of the change of the $Z_2$ topological number and associated surface states, 
the evolution of the surface states accompanying the WTI-STI topological phase transition occurs 
in the following way. When the I-symmetry is broken, there should generally arise a WSM phase between the 
WTI-STI phase transition.   
In the WTI-WSM phase transition, two pairs of Weyl nodes are created close to a TRIM $\bm{k}_{\Gamma}$
\cite{Murakami07b}. As 
the system enters the WSM phase, 
a Fermi arc is formed between the 
two Weyl nodes within each pair. Thus there are two Fermi arcs which are symmetric with respect to the TRIM. 
As a control parameter is changed, the Fermi arcs grow as the Weyl nodes travel around the TRIM. 
Eventually, at the WSM-STI phase transition, the four Weyl nodes annihilate pairwise, causing a fusion of
two Fermi arcs into a single Dirac cone encircling the TRIM, as shown by Fig.~\ref{arcandcone} .

\begin{figure}[t]
\includegraphics[width=7.5cm]{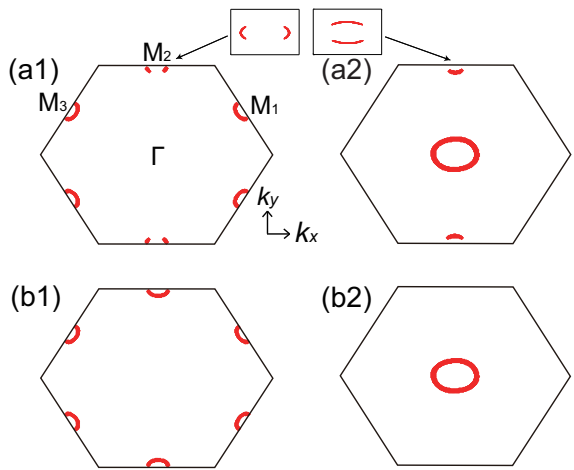}
\caption{\label{dbnew}The surface Fermi surfaces at $E=0$ in the whole BZ when $\lambda _v=0.2t$.
The values of $\delta t_+$ is $\delta t_+ =0$ (WSM) for (a1) and (a2), and $\delta t_+=0.05t$ (STI) for (b1) and (b2).
In (a1) and (b1), the surfaces are terminated without dangling bonds, and in (a2) and (b2) with dangling bonds.
The insets show the magnified images of the surface Fermi surface around the $M_2$ point. 
}
\end{figure}

In Ref.~\onlinecite{Teo}, it is argued that when I symmetry is preserved, a $Z_2$ topological index defined for each surface TRIM indicates 
whether the focused surface TRIM is inside or outside the surface Fermi surface. It is also concluded that this index depends on 
the surface termination.
Within this argument in Ref.~\onlinecite{Teo}, the surface should include the inversion center, and therefore 
there are two possible surface terminations for a fixed surface orientation. 
If we change one surface termination into the other, surface TRIM which were inside the Fermi surface will become outside the 
Fermi surface, and vice versa. We now try to apply this scenario to our model. However, 
the I symmetry is broken in our model, and the discussion in Ref.~\onlinecite{Teo} is not directly applied, 
Nevertheless, we can expect the similar physics from continuity argument, by switching on the I-symmetry breaking. 
For example, in Figs.~\ref{dbnew} (a1) and (b1), we show the Fermi surface on the (111) surface with the surface terminated with the atoms, each of which has three bonds
along 1$\bar{1}\bar{1}$, $\bar{1}1\bar{1}$, and $\bar{1}\bar{1}1$. In this surface termination, the top surface
is terminated by atoms in the A sublattice, and the bottom surface by atoms in the B sublattice. 
By adding bonds (i.e., ``dangling bonds") along 111 directions to the topmost atoms, we can switch from one 
surface termination to the other, namely the top and the bottom surfaces terminated by B and A sublattices, respectively. 
The results are plotted in 
  Figs.~\ref{dbnew} (a2) and (b2), whose parameters are identical with (a1) and (b1), respectively. We can see that the physics discussed in Ref.~\onlinecite{Teo}
  carries over to the present model as well. For example, the M$_1$ and M$_3$ points are inside the Fermi surfaces 
  when the dangling bonds are absent [Figs.~\ref{dbnew}(a1) and (b1)], but when the dangling bonds are added, the Fermi surfaces around 
 the M$_1$ and M$_3$ points disappear [Figs.~\ref{dbnew}(a2) and (b2)]. On the other hand, there appear a new Fermi surface around the $\Gamma$ point
 when the dangling bonds are added. The remarkable phenomenon occurs around the M$_2$ point. The 
Fermi surface around the M$_2$ point in the STI phase in (b1) disappears in the plot (b2) where the dangling bonds are
present. This also affects the neighboring WSM phase, as can be seen by comparing (a1) and (a2). Among the Weyl nodes in (a1) 
the Fermi arcs arise between W$_{1+}$-W$_{1-}$ and between W$_{2+}$-W$_{2-}$. Meanwhile in (a2), 
the Fermi arcs arise between W$_{1+}$-W$_{2-}$ and between W$_{2+}$-W$_{1-}$. Thus we have shown that the change of 
surface termination exchanges the pairs of Weyl nodes, out of which the Fermi arcs are formed. 

{This change of pairing of 
Weyl nodes by varying surface terminations occurs in generic WSMs. 
The Dirac cones in TIs depend on surface terminations, as shown in Ref.~\onlinecite{Teo}.
Because 
the WSM phase is next to the TI phase \cite{Murakami07b,MurakamiKuga}, the dependence on the surface termination in general WSMs 
(with TR symmetry)
follows from that in the TIs, as we discussed in this paper. 
When the surface termination is varied, the pairing of the Weyl nodes will change, and 
the union of the pairing of the Weyl nodes before and after the change of surface termination 
forms a loop, which turns out to be the surface Fermi surface in the TI phase 
around a particular TRIM.  In the present case, the pairing is 
$\{(\mathrm{W}_{1+},\mathrm{W}_{1-}), (\mathrm{W}_{2+},\mathrm{W}_{2-})\}$
or $\{(\mathrm{W}_{1+},\mathrm{W}_{2-}), (\mathrm{W}_{2+},\mathrm{W}_{1-})\}$, depending on the
surface termination, and 
their union forms a loop around the M$_2$ point.
 This also implies that the pairing of the Weyl nodes for the
Fermi arc is not solely determined from bulk band structure, because it depends on surface terminations.  }


\section{CONCLUSION}
In the present paper, we study dispersions of Fermi arcs in the Weyl semimetal phase. We first construct a 
simple effective model, describing the Weyl semimetal with two Weyl nodes close to each other. 
We find that the dispersions of Fermi-arc states 
for top- and bottom surfaces cross around the Weyl point, and they have opposite velocities. These Fermi-arc dispersions 
are tangential to the bulk Dirac cones around the Weyl points. 
These results are confirmed by a calculation using a tight-binding model with time-reversal symmetry but without inversion symmetry. 
In this model calculations, we see 
that the Fermi arcs gradually grow by changing a model parameter, and that two Fermi arcs finally 
merge together to form a single Dirac cone when the system transits from the Weyl semimetal to the topological insulator phase.  
We also find that by changing the surface termination, the pairing between the two monopoles and two antimonopoles 
to make Fermi arcs is switched. These results reveal an interesting interplay between the surface and the bulk 
electronic states in Weyl semimetals and topological insulators.

\begin{acknowledgments}
We are grateful to L. Balents and F. D. M. Haldane
for fruitful discussions.
This research is supported in part 
by MEXT KAKENHI Grant Nos. 22540327 and 26287062, by MEXT Elements Strategy Initiative to Form Core Research Center (TIES),
 and by the National Science Foundation under
Grant No.~NSF PHY11-25915 through Kavli Institute for Theoretical Physics, University of California at
Santa Barbara, where part of the present work was completed.  

\end{acknowledgments}

\appendix
\section{Effective model close to monopole-antimonopole pair creation or annihilation 
in $\bm{k}$ space}
\label{sec:app-A}

From the Hamiltonian (\ref{eq:hamiltonian}) with (\ref{eq:a-matrix}), one can derive 
an effective model describing the WSM phase close to a monopole-antimonopole pair creation/annihilation.
The argument closely follows that in Ref.~\onlinecite{MurakamiKuga}.
We note that $\Delta m$ is a control parameter for the Hamiltonian, and our goal is to 
construct a Hamiltonian where positive and negative $\Delta m$
represents the WSM and the bulk insulating phases, respectively. 
First we note that the determinant of the matrix $M$ in (\ref{eq:a-matrix}) is zero, because 
otherwise the gapless condition (\ref{eq:gaplesscondition}) guarantees existence of $\bm{k}$ for both the positive 
and the negative values of $\Delta m$, meaning that both the positive and negative sides of $\Delta m$ are conducting
in the bulk.
Hence we have ${\rm det} M=0$, and therefore, the matrix $M$ has a unit eigenvector
$\bm{n}_{1}$ with
zero eigenvalue: $M\bm{n}_{1}=0$.
In Ref.~\onlinecite{MurakamiKuga}, an orthonormal basis $\left\{\bm{n}_{1},\bm{n}_{2},\bm{n}_{3}
\right\}$ is constructed out of this unit vector $\bm{n}_1$.
While we can in principle proceed here as in Ref.~\onlinecite{MurakamiKuga},
it leaves a number of free parameters in the model. In fact we can always set
$\bm{n}_1={}^{t}(1,0,0)$, $\bm{n}_2={}^{t}(0,1,0)$, 
$\bm{n}_3={}^{t}(0,0,1)$, by a rotation of $\bm{k}$ coordinate axes. 
Then from (\ref{eq:a-matrix}) to the linear order in 
$\Delta\bm{k}$ and $m$,
we have
\begin{equation}
\bm{a}=\Delta k_{y}\bm{u}_{2}+
\Delta k_{z}\bm{u}_{3}+\Delta m \bm{N},
\end{equation}
where $\bm{u}_{i}=M\bm{n}_{i}$ ($i=2,3$). 
The gap closes when $\bm{a}=0$, but it cannot happen in general for $\Delta m\neq 0$ 
because the three vectors $\bm{u}_{2}$, 
$\bm{u}_{3}$, 
$\bm{N}$ are generally linearly independent.
It is an artifact of retaining only the linear terms in $\Delta m$ and $\Delta \bm{k}$. 
Thus we have to include the next order in $\Delta \bm{k}$ and
$\Delta m$. It turns out that the only relevant term here is the quadratic term in $\Delta k_x$ 
\cite{MurakamiKuga}
and therefore we additionally retain only this term, to obtain
\begin{equation}
\bm{a}=
\Delta m \bm{N}+\Delta k_{y}\bm{u}_{2}+
\Delta k_{z}\bm{u}_{3}+
(\Delta k_{x})^{2}\bm{u}_{11},
\end{equation}
where $\gamma$ and $v$ are positive constants. 
The gap closes when this vector is zero. The solution can be written down explicitly for generic cases, 
but instead we here introduce a simplifying assumption
\begin{equation}
\bm{u}_{11}=-\bm{N}=\left(
\begin{array}{c}
\gamma\\0\\0
\end{array}\right),\ 
\bm{u}_2=\left(
\begin{array}{c}
0\\v\\0
\end{array}\right),
\bm{u}_3=\left(
\begin{array}{c}
0\\0\\v
\end{array}\right).
\label{eq:simplifying}
\end{equation}
Namely, $\bm{u}_{11}$ and $\bm{N}$ are parallel, and they are orthogonal to 
$\bm{u}_2$ and $\bm{u}_3$. Then we get
\begin{equation}
\bm{a}=\left(
\begin{array}{c}
\gamma((\Delta p_{1})^{2}-m) \\ v\Delta p_{2}\\ v\Delta p_{3}
\end{array}\right).
\end{equation}
It is straightforward to see that 
the solution for this exists only when $m$ is positive. This means that positive $m$ represents the 
WSM phase while negative $m$ means a bulk insulating phase. Within the WSM phase, the Weyl nodes 
are given by $(\Delta k_{x},\Delta k_{y},\Delta k_{z})=(\pm \sqrt{m},0,0)$.
If we have not employed the simplifying assumptions Eq.~(\ref{eq:simplifying}), there  arise terms in the expressions 
of the Weyl nodes, which are linear in $m$ \cite{MurakamiKuga}. Nevertheless, these terms are not the
main focus of the paper, and we discard them for simplicity for illustration of surface state dispersions and evolutions.

\bibliographystyle{apsrev4-1}
\bibliography{ronbunws2}

\end{document}